\documentclass[10pt,conference]{IEEEtran}
\IEEEoverridecommandlockouts

\usepackage[T1]{fontenc}
\usepackage[utf8]{inputenc}
\usepackage[british]{babel}
\usepackage{microtype}
\usepackage{booktabs}
\usepackage{tabularx}
\usepackage{graphicx}
\usepackage{amsmath}
\usepackage{xcolor}
\usepackage{url}
\usepackage{tikz}
\usepackage{dblfloatfix}
\usetikzlibrary{arrows.meta, positioning, fit, calc, backgrounds}
\usepackage[hidelinks]{hyperref}
\usepackage{balance}
\usepackage{listings}
\lstdefinestyle{skillfile}{
  basicstyle=\ttfamily\scriptsize,
  columns=fullflexible,
  keepspaces=true,
  breaklines=true,
  breakindent=1.2em,
  frame=single,
  framerule=0.4pt,
  rulecolor=\color{black!60},
  xleftmargin=2pt, xrightmargin=2pt,
  aboveskip=2pt, belowskip=2pt,
  morekeywords={name,description,license},
  keywordstyle=\color{arblue},
  morecomment=[l][\color{arblue}\bfseries]\#,
  moredelim=[l][\color{ingray}]{---},
  moredelim=[s][\color{accgreen}]{`}{`},
}

\definecolor{arblue}{HTML}{3730A3}
\definecolor{ingray}{HTML}{6B7280}
\definecolor{accgreen}{HTML}{059669}
\definecolor{stubred}{HTML}{B91C1C}
\tikzset{
  srcnode/.style={rounded corners=4pt, draw=ingray!70, fill=ingray!10,
                  text width=1.5cm, minimum height=2.4cm, align=center,
                  font=\scriptsize, inner sep=4pt},
  stagenode/.style={rounded corners=4pt, draw=arblue!70, fill=arblue!6,
                    line width=0.7pt, text width=2.95cm, minimum height=2.4cm,
                    align=left, font=\scriptsize, inner sep=5pt},
  storenode/.style={rounded corners=4pt, draw=accgreen!70, fill=accgreen!8,
                    line width=0.7pt, text width=2.5cm, minimum height=2.4cm,
                    align=left, font=\scriptsize, inner sep=5pt},
  pipeflow/.style={-{Stealth}, thick, arblue!75},
  fbar/.style={draw=arblue!70, fill=arblue!12, minimum height=7.5mm,
               font=\scriptsize, align=center, rounded corners=2pt},
  fkeep/.style={-{Stealth}, semithick, arblue!75},
  fdrop/.style={font=\scriptsize\itshape, text=stubred, align=left},
}
\newcommand{\stagehead}[1]{{\bfseries\color{arblue}#1}}
\newcommand{\storehead}[1]{{\bfseries\color{accgreen!60!black}#1}}

\newcommand{\datasetname}{GitSkills}          %
\newcommand{\dataDate}{July 2026}             %
\newcommand{\numSearchEstimate}{349{,}000}    %
\newcommand{\numFilesTotal}{3{,}797{,}117}    %
\newcommand{\numUnique}{1{,}877{,}981}        %
\newcommand{\numRepos}{282{,}200}             %
\newcommand{\numOwners}{195{,}841}            %
\newcommand{\dupShare}{50.5\%}                %
\newcommand{\numSiblings}{7{,}264{,}865}      %
\newcommand{\numCommitsSampled}{458{,}548}    %
\newcommand{\formatLaunch}{October 2025}         %
\newcommand{\numFilesApprox}{3.8M}  %
\newcommand{\rqcat}[1]{\smallskip\noindent\textbf{#1}}
\usepackage{fancyhdr}

\fancypagestyle{citethis}{
  \fancyhf{} %
  
  \fancyfoot[C]{%
    \begin{minipage}{\textwidth}
      \rule{\textwidth}{0.4pt}\\[3pt]
      \scriptsize\raggedright
      \textbf{If you use the \datasetname \ dataset, cite this paper as:}\\
      G. Destefanis, D. Graziotin, M. Vaccargiu, and M. Ortu, ``\datasetname: A Dataset of Agent Skills on GitHub,'' in \emph{Proceedings of the 24th IEEE/ACM International Conference on Mining Software Repositories (MSR '27)}. IEEE, Piscataway, NJ, USA, 2027, 3 pp. To appear.
    \end{minipage}%
  }
}
\begin{document}

\title{\datasetname: A Dataset of Agent Skills on GitHub}

\author{
\IEEEauthorblockN{Giuseppe Destefanis\IEEEauthorrefmark{1},
Daniel Graziotin\IEEEauthorrefmark{2},
Matteo Vaccargiu\IEEEauthorrefmark{2},
Marco Ortu\IEEEauthorrefmark{3}}
\IEEEauthorblockA{\IEEEauthorrefmark{1}University College London, United Kingdom
\quad g.destefanis@ucl.ac.uk}
\IEEEauthorblockA{\IEEEauthorrefmark{2}University of Hohenheim, Stuttgart, Germany
\quad \{graziotin, matteo.vaccargiu\}@uni-hohenheim.de}
\IEEEauthorblockA{\IEEEauthorrefmark{3}University of Cagliari, Italy
\quad marco.ortu@unica.it}
}
\maketitle
\thispagestyle{citethis}
\begin{abstract}
An agent skill is a folder containing a \texttt{SKILL.md} file with
instructions for a language-model agent, optionally accompanied by
scripts and reference files. The agent loads the
skill when it judges that a task matches the skill description.
Anthropic introduced the format in \formatLaunch{} as an open specification.
Nine months
later, public GitHub repositories hold millions of skill files. Skills are unlike the artifacts that software engineering researchers usually
mine: they are written mainly in natural language, a
model selects them probabilistically at run time, and no compiler or type checker verifies the selection. Skills also have no central registry or
package manager; developers reuse them by copying folders between
repositories. How
developers write, reuse, and maintain skills is therefore an
empirical question, and no existing dataset records this population.  We present
\datasetname, a dataset of \numFilesTotal{} \texttt{SKILL.md} files collected
from \numRepos{} public repositories in \dataDate. The dataset retains every
file occurrence with its repository, path, and content hash. We group
identical files into \numUnique{} distinct contents and enrich one
representative per group with the full text, parsed front matter,
folder contents, repository metadata, and, for a subset, the commit history of the file. A single self-contained SQLite file supports research on the adoption,
reuse, structure, authorship, maintenance, and security of agent skills.
\end{abstract}


\section{High-Level Overview}

An agent skill describes a workflow, convention, or procedure that a
language-model agent should follow for a class of tasks. A skill consists of
a folder containing a \texttt{SKILL.md} file. The file has YAML front matter
with a name and description, followed by a Markdown body with the
instructions (Figure~\ref{fig:skillexample} in the Appendix). The folder may also contain scripts and reference documents.
Anthropic introduced the format in \formatLaunch{} and published its
directory structure, front-matter constraints, and staged loading model as an
open specification~\cite{agentskills,anthropicskills}. Any agent tool can
support the format. Claude Code is the reference
implementation~\cite{claudecodeskills}; where a tool-specific
convention matters, such as the \texttt{.claude/skills/} skill
directory, we follow Claude Code's for simplicity.

The model selects skills probabilistically: at run time it compares the task
with the skill description and decides whether to load the skill, a step
that no compiler or type checker verifies. A vague
description may prevent selection, while unclear instructions may lead to
incomplete or incorrect execution without an explicit error. These problems
are difficult to detect with conventional software analysis. How developers
write and maintain skills is therefore an empirical question, but no dataset
currently supports a large-scale study of this artifact.

\datasetname{} records \numFilesTotal{}
\texttt{SKILL.md} files from \numRepos{} public GitHub repositories owned by
\numOwners{} accounts, collected in \dataDate. We group the files by
content hash into \numUnique{} distinct contents and enrich one
representative file per group with its full text, parsed front matter, folder
contents, and repository metadata; commit history, with first- and
last-commit author accounts, covers skills in standard locations
(recognized skill directories such as \texttt{.claude/skills/}) and
a size-stratified sample of the rest. Every occurrence is retained with its
repository and path, so researchers can study both unique contents and
their copies.

The dataset covers the early adoption of the format, while conventions and
tooling are still developing, so researchers can observe how a new
software artifact spreads, how common practices emerge, and whether the
format develops into shared infrastructure across agent tools.

We collected the dataset with a read-only pipeline against the GitHub
code-search and REST APIs. Because code search returns at most 1{,}000
results per query, discovery partitions the search space by file size
until every range can be retrieved completely.
Appendix~\ref{sec:collection} details the pipeline and its coverage
limits; the dataset covers public repositories only and is a lower
bound on the population.

\section{Internal Structure}

Table~\ref{tab:structure} summarizes the dataset tables. The dataset is stored in SQLite format.

When several repositories contain the same \texttt{SKILL.md} file, the dataset stores the text once and links all copies to it through the content hash. Identical hashes mean that the files contain exactly the same bytes.

We collect folder contents and commit history for only one copy of each distinct skill. This information applies only to that specific repository. Other copies may have different scripts, reference files, or commit histories.

\begin{table*}[t]
\caption{Overview of the \datasetname{} dataset.}
\label{tab:structure}
\centering
\begin{tabularx}{\textwidth}{llrX}
\toprule
& Table & \# Records & Content \\
\midrule
Core &
\texttt{artifacts} & \numFilesTotal &
One row per discovered file: repository, path, exact basename, location
class, content hash, and representative flag; for representatives, also the
full text, parsed front matter, and body size. \\
& \texttt{repos} & \numRepos &
Repository metadata: owner, star count, primary language, fork status,
creation date, and last-push date. \\
\midrule
Composition &
\texttt{artifact\_siblings} & \numSiblings &
Files stored alongside a representative skill: path, entry type (file or directory), size, and the text of files under a size cap. Whether a skill bundles scripts or reference material is recorded per skill in artifacts (has\_scripts, has\_references).\\
\midrule
History &
(columns in \texttt{artifacts}) & \numCommitsSampled &
First and last commit dates of the \texttt{SKILL.md} file, their author
accounts (anonymized; user or bot), and commit count, for standard
locations and a size-stratified sample of the rest. \\
\midrule
Collection &
\texttt{mining\_runs} & 7 &
Query, start and end timestamps, and search-result count for each collection run; four runs that crashed were closed retroactively with a zero count. \\
\bottomrule
\end{tabularx}
\end{table*}

\begin{figure*}[!t]
\centering
\begin{tikzpicture}[node distance=7mm]
\node[srcnode] (gh) {\textbf{GitHub}\\[2pt]public\\repositories};
\node[stagenode, right=of gh] (disc)
  {\stagehead{1. Discovery}\\[3pt]
   search for files matching \texttt{SKILL.md};
   partition by file size to pass the 1,000-result limit;
   classify each result by location\\[2pt]
   \emph{retain every result}};
\node[stagenode, right=of disc] (dedup)
  {\stagehead{2. Deduplication}\\[3pt]
   group files by content hash; select one representative per group,
   preferring a standard location\\[2pt]
   \emph{retain all copies}};
\node[stagenode, right=of dedup] (enr)
  {\stagehead{3. Enrichment}\\[3pt]
   \emph{for representatives:}
   content and front matter; folder contents; repository metadata;
   sampled \texttt{SKILL.md} commit history};
\node[storenode, right=of enr] (db)
  {\storehead{Dataset (SQLite)}\\[3pt]
   \texttt{artifacts}\\
   \texttt{artifact\_siblings}\\
   \texttt{repos}\\
   \texttt{mining\_runs}};
\draw[pipeflow] (gh) -- (disc);
\draw[pipeflow] (disc) -- (dedup);
\draw[pipeflow] (dedup) -- (enr);
\draw[pipeflow] (enr) -- (db);
\end{tikzpicture}
\caption{Collection pipeline. Discovery retains every filename match; deduplication selects one representative per distinct content while retaining all copies; enrichment applies to the representatives.}
\label{fig:pipeline}
\end{figure*}
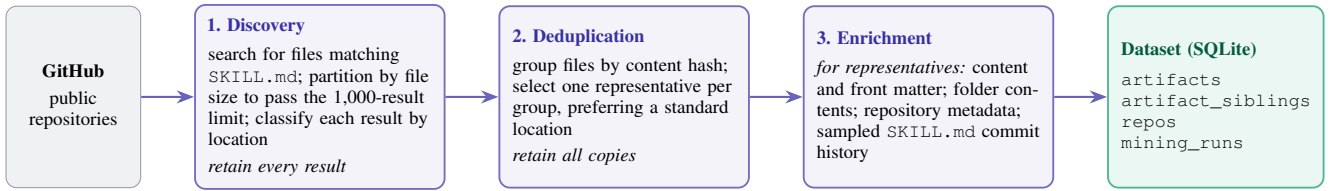

\section{Potential Research Questions}
Agent skills influence how coding agents operate within software
projects, so their content, structure, reuse, and maintenance are
empirical questions, which researchers can now study at the level of the population.

\rqcat{1) Adoption and linguistic evolution.}
\begin{itemize}
\item[a)] How quickly does the format spread, and which projects adopt
it first, in terms of programming language, popularity, age, and
activity?
\item[b)] What do developers codify in skills, and in which contexts do
skills appear, from operational projects to catalogs, templates, and
demonstrations? A taxonomy of skill purposes does not yet exist.
\item[c)] Do the linguistic properties of newly written skills change
across monthly cohorts, in structure and phrasing as well as in topic
coverage and semantic diversity? Convergence toward formulaic templates
would indicate an emerging genre; shrinking diversity may also reflect
rising machine authorship.
\item[d)] How many skills do agents themselves create or maintain, and
in which natural languages are skills written? A skill is read by a
multilingual model, so a developer may state a procedure more precisely
in their own language than in English.
\end{itemize}

\rqcat{2) Development of a shared format.}
\begin{itemize}
\item[a)] What proportion of skills use vendor-neutral rather than
tool-specific locations, and how does this proportion change over time?
\item[b)] Do skill texts address one named tool, or any agent that
implements the specification?
\end{itemize}

\rqcat{3) Reuse without a package manager.} Skills have no central
registry; reuse happens by copying folders, and \dupShare{} of the
collected files are byte-for-byte copies of another file in the dataset.
\begin{itemize}
\item[a)] How concentrated is reuse: a long tail of rarely copied
contents, or a small set of widely copied templates?
\item[b)] Through which mechanisms do skills move between repositories,
such as direct addition, catalogs, or scaffolding tools?
\item[c)] Do skill copies follow the genealogy patterns known from code
clones, such as consistent and inconsistent propagation of changes?
\end{itemize}

\rqcat{4) Software metrics for natural-language instructions.}
\begin{itemize}
\item[a)] Which established metrics, such as size, churn, age, clone
coverage, and readability, have meaningful equivalents for skills, and
how do their distributions compare with those of source code?
\item[b)] Can observable indicators of skill quality be defined and
compared with proxies such as copy count and subsequent edits?
\item[c)] Which properties of the description, the text the agent
matches against when deciding whether to load the skill, are associated
with reuse and maintenance?
\end{itemize}

\rqcat{5) Maintenance and trust.} Skills can instruct agents to run
commands, access external resources, and execute bundled scripts, and
they are copied between repositories without formal review.
\begin{itemize}
\item[a)] How often do skills become outdated relative to the projects
and tools they describe?
\item[b)] Do modified copies of widely reused skills introduce command
execution or network access absent from the original, the analog of a
supply-chain attack in an ecosystem without a registry?
\item[c)] How often do skills bundle executable files, and how widely
are these skills copied?
\end{itemize}

\section{How to Access}

The full dataset, as a single self-contained SQLite file, is archived on
Zenodo at this
\href{https://doi.org/10.5281/zenodo.21875637}{link}\footnote{\url{https://doi.org/10.5281/zenodo.21875637}}.
A Parquet mirror, partitioned by table, is available on Hugging Face at
this
\href{https://huggingface.co/datasets/mvaccargiu/gitskills}{link}\footnote{\url{https://huggingface.co/datasets/mvaccargiu/gitskills}}.

A sample of the dataset is available on GitHub at this \href{https://github.com/giuseppedestefanis/gitskills-sample}{link}\footnote{\url{https://github.com/giuseppedestefanis/gitskills-sample}}.

We replaced commit author accounts with keyed one-way codes, identical
for the same account throughout, so authorship can be traced without
identifying anyone. Bot accounts keep
their login. We redacted email addresses and personal names in commit
messages and kept AI assistant names in \texttt{Co-authored-by}
trailers.

\smallskip\noindent\textbf{Acknowledgement.} Matteo Vaccargiu has been supported by the Hector Stiftung.

\appendices

\section{Dataset Construction}
\label{sec:collection}

Figure~\ref{fig:pipeline} shows the collection pipeline. Collection is read-only; requests go to the GitHub REST and GraphQL APIs, the code-search API, and the raw-content CDN. Each stage checkpoints its progress in the database and resumes after interruptions. 

\emph{Discovery.} Agent tools identify skills by filename, so the exact basename is a direct marker for code search. The code-search API returns at most 1{,}000 results per query, and its \texttt{total\_count} estimate proved unreliable: it reported roughly \numSearchEstimate{} matches for the filename query, against over \numFilesApprox{} files ultimately retrieved. We therefore partitioned the search space by file size, splitting any range with more than 1{,}000 results until every range
could be retrieved completely.

\emph{Deduplication.} We group files by content hash and select one representative per group for enrichment, preferring a file in the \texttt{.claude/skills/} directory, with a deterministic rule to
break ties. We do not assume that the representative is the original source. All copies remain in the dataset with their repository, path, and location class, so the spread of each content can be measured.

\emph{Enrichment.} For each representative, the tool downloads the \texttt{SKILL.md} file, parses its front matter, and records the bundled scripts and reference documents in the skill folder, downloading the text of bundled files up to a size cap. It also collects repository metadata, including star counts, and retrieves the commit history of the \texttt{SKILL.md} file, with the author account of the first and last commit stored as an anonymized code, for skills in standard locations and a size-stratified sample of the others; the file's own history dates the skill's addition. Folder listings are missing for 1{,}544 representatives and commit history for 1{,}166 of those in standard locations. For 42 representatives the file changed on GitHub between discovery and fetch and the original could not be recovered; \texttt{content\_sha\_ok} records this.

\emph{Anonymization.} We masked email addresses in commit messages,
including GitHub noreply addresses, with a fixed marker, and a scan
of the released file confirmed that none remain; the first and last
commit messages are included in this redacted form. We replaced commit
author accounts and personal names in trailer lines with codes
from a keyed one-way function; the codes cannot be reversed and stay
stable across the dataset.

The search matches the basename case-insensitively and also returns case variants such as \texttt{skill.md} (54{,}289 files, 1.4\%), including lowercase files predating the format. We retained them: each record carries the exact basename,
location class, front-matter validity, and date of the first recorded commit, so researchers can define and compare stricter inclusion criteria during analysis. The dataset covers public repositories only, and GitHub code search indexes only default branches, files under 384~KB, recently active repositories with fewer than 500{,}000 files, and forks only when they have more stars than the parent repository. The dataset is therefore a lower bound on the full population.

\begin{figure}[!t]
\begin{lstlisting}[style=skillfile]
---
name: web-artifacts-builder
description: Suite of tools for creating elaborate, multi-component claude.ai HTML artifacts using modern frontend web technologies (React, Tailwind CSS, shadcn/ui). Use for complex artifacts requiring state management, routing, or shadcn/ui components - not for simple single-file HTML/JSX artifacts.
license: Complete terms in LICENSE.txt
---

# Web Artifacts Builder

To build powerful frontend claude.ai artifacts, follow these steps:
1. Initialize the frontend repo using `scripts/init-artifact.sh`
2. Develop your artifact by editing the generated code
3. Bundle all code into a single HTML file using `scripts/bundle-artifact.sh`
4. Display artifact to user
5. (Optional) Test the artifact

[...]
\end{lstlisting}
\caption{Opening of the \texttt{SKILL.md} of Anthropic's
\texttt{web-artifacts-builder} skill, reproduced verbatim and
truncated. Source:
\url{https://github.com/anthropics/skills}. The YAML front matter
carries the name and the description that the agent matches against
the task when deciding whether to load the skill; the Markdown body
holds the instructions, here referencing bundled executable scripts.}
\label{fig:skillexample}
\end{figure}

\end{document}